Astrophysical Wake Acceleration Driven by Relativistic Alfvenic Pulse Emitted from Bursting Accretion Disk

Toshikazu Ebisuzaki[1] and Toshiki Tajima[2]

[1]RIKEN Cluster for Pioneering Research, 2-1 Hirosawa, Wako 351-1198, Japan, ebisu@postman.riken.jp

[2]Depart. of Physics and Astronmy, University of California, Irvine, CA 92679, USA, ttajima@uci.edu

Abstract

We consider that electromagnetic pulses produced in the jets of this innermost part of the accretion disk accelerate charged particles (protons, ions, electrons) to very high energies including energies above $10^{20}$ eV for the case of protons and nucleus and $10^{12-15}$ eV for electrons by electromagnetic wave-particle interaction. The episodic eruptive accretion in the disk by the magneto-rotational instability gives rise to the strong Alfvenic pulses, which acts as the driver of the collective accelerating pondermotive force. This pondermotive force drives the wakes. The accelerated hadrons (protons and nuclei) are released to the intergalactic space to be ultra-high energy cosmic rays. The high-energy electrons, on the other hand, emit photons in the collisions of electromagnetic perturbances to produce various non-thermal emissions (radio, IR, visible, UV, and gamma-rays) of active galactic nuclei. Applying the theory to M82 X-1, we find that it can explain the northern hot spot of ultra high energy cosmic rays above $6 \times 10^{19}$ eV. We also discuss astrophysical implications for other nearby active galactic nuclei, neutron star mergers, and high energy neutrinos.

1. Introduction

The Event Horizon Telescope (EHT) Collaboration released the first direct image of the central blackhole of M87 galaxy synthesized by the observational data taken by many radio telescopes around the Earth by VLBI technology (Figure 1; [1-6]). According to them, the image captures the innermost part of the accretion disk distorted by strong gravitational field of the blackhole (blackhole shadow). The surprising similarity of the observed image of blackhole to those of theoretical works (e.g. Tajima and Shibata [7]) confirms the validity of the accreting blackhole model of the active galactic nuclei, such as M87.

In the present paper, we consider that electromagnetic pulses produced in the jets of the innermost part of the accretion disk accelerate charged particles (protons, ions, electrons) to very high energies including energies above $10^{20}$ eV for the case of protons and nucleus and $10^{12-15}$ eV for electrons by electromagnetic wave-particle interaction. The episodic eruptive accretion in the disk by the magneto-rotational instability [8,9] gives rise to the strong (superrelativistic, i.e., the electron momentum far exceeding $m_e c$, where $m_e$ is the mass of electrons) Alfvenic pulses, which acts as the driver of the collective accelerating pondermotive force [10]. This pondermotive force drives the wakes (both bow and stern wakes [11,12]). As investigated in the present super-relativistic pulse case, the bow wake far exceeds the conventional (stern) wake acceleration (see the comparison of the bow and stern wakes in s.a. [13]). The accelerated hadrons (protons and nuclei) are released to the intergalactic space to be ultra-high energy cosmic rays. The high-energy electrons, on the other hand, emit photons in the collisions of electromagnetic perturbances to produce various non-thermal emissions (radio, IR, visible, UV, and gamma-rays) of active galactic nuclei.

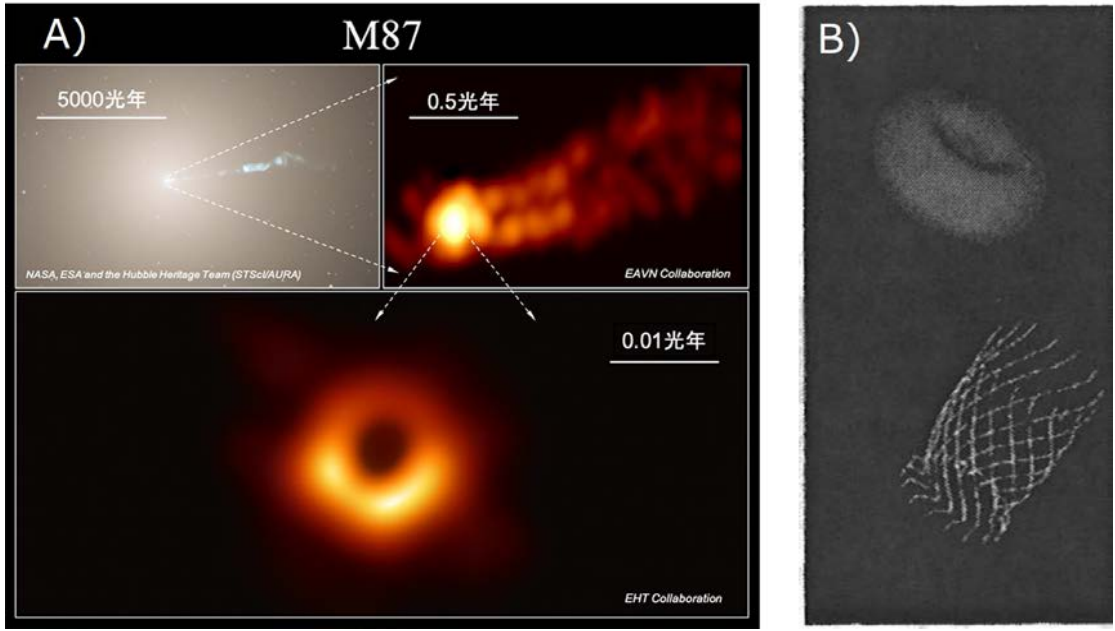

Figure 1. A) Image of the central blackhole of M87 synthesized by the Event Horizon Telescope collaboration and that of jets by East Asia VLBI network. (EHT collaboration ([1-6]; https://www.nao.ac.jp/news/sp/20190410-eht/images.html). B) Artistic picture of an accretion disk around a blackhole, taken from Tajima and Shibata [7].

The conditions for this strong acceleration by wakefield are:

(a) the acceleration structure (wave) is very close to the relativistic propagation velocity (phase velocity), i.e. the speed of light; and

(b) the wave has a relativistic amplitude (i.e. the particles in the wave have a relativistic momentum $e_j E/\omega > m_j c$ in one photon cycle). Where $e_j$ and $m_j$ are the charge and the mass of the particle $j$, and $E$ and $\omega$ are the electric field and angular frequency of the wave.

The condition (b) came from the fact that the significant acceleration by a pondermotive force of the electromagnetic wave takes place only in the nonlinear case: This term becomes significant only if the amplitudes of the electromagnetic waves become relativistic [10,14,15]. If these two conditions are satisfied, as shown by a number of ground experiments, [16], this acceleration mechanism is superior to Fermi acceleration [17], in which charged particles gradually gain energy as they are scattered many times by magnetic clouds. for the following reasons.

1) The pondermotive force provides a very high acceleration field.
2) No deflection of the particles is required which causes energy loss due to synchrotron radiation.
3) The acceleration field is parallel to the direction of movement of the particles and has the same velocity (speed of light). The result is a robust built-in coherence in acceleration systems, which is called as relativistic coherence [18]. In such cases, the energy spectrum takes the form of a power function of the exponent -2 ($E^{-2}$; see [19,20]) due to the phase difference between the particle and the acceleration field. On the other hand, Fermi acceleration based on a large number of scattering is incoherent, it is an extended, and has no specific temporal structure.
4) There are no escape problems [21].
5) The acceleration field dissipates spontaneously so that the particles are free to escape from the field after the acceleration. The wave has such coherent dynamics as long as it is excited, with a sufficiently high frequency. On the other hand, in a mechanism that accelerates with an electrostatic structure [22], the acceleration of the charged particles is not possible unless there is a special reason to overcome the plasma and maintain the structure since the plasma itself has a strong tendency to destroy the electrostatic field. Also, there is no reason to have a low spectrum power.

This new acceleration mechanism, therefore, seams to solve a long-standing enigma of the origin of ultra-high energy cosmic rays (Ultra-High Energy Cosmic rays: UHECR) with energy of $10^{20}$ eV (e.g., [21]), in which have been discussed primarily in the framework of the Fermi acceleration [17]. The candidate astronomical objects of ULHECR were neutron stars, active galactic nuclei, gamma-ray bursts, and cosmological accretion shock waves in the intergalactic space. However, even in these candidates, the acceleration to $10^{20}$ eV by the Fermi mechanism was difficult because of : (1) very large number of scatterings is required to reach high energy; (2) energy loss due to the synchrotron radiation is not negligible at the time of scattering; and (3) adiabatic energy loss takes place when particles escape from the acceleration region.

    The idea that electromagnetic waves accelerate charged particles is not new. For example, Takahashi et al. [23] and Chen et al. [20] showed that strong Alfven waves, produced by a neutron star collision were able to accelerate charged particles to energies above $10^{20}$ eV. Although it is believed to be associated with such short gamma bursts [24], the direct collisions of the two neutron stars are not very likely. This requires exactly the same masses, for the binary neutron stars because otherwise the more massive neutron star destroys the lighter stars into an accretion disk. Chang et al. [25] performed one-dimensional numerical calculations of the propagation of whistler wave emitted from AGNs and found that a wakefield accelerates particles to UHECR.

    Accreting supermassive blackhole is the main engine of AGNs, in which Ebisuzaki and Tajima [11,12] considered wakefield acceleration to take place. Accretion disk has been shown to repeated transitions between a strongly magnetized (low ß) state and a weakly magnetized (high ß) state [26]. In fact, O'Neil et al. [27] shows that the transition takes place every 10-20 orbital periods in the three-dimensional simulation. The amplitude of the distortion in the magnetic field becomes resulting in a very large amplitude at the innermost portion of the disk. At this transition from the strongly magnetized state to the weakly magnetized state, strong pulses of electromagnetic disturbance are excited in the accretion disk. This disturbance converted into strong pondermotive field by nonlinear effects in jets made of plasmas ejected from accretion disk with relativistic velocities. It is shown that this pondermotive force can spontaneously accelerate protons and nuclei to ultimate energies exceeding ZeV ($10^{21}$ eV). Mizuta et al. [28] performed three-dimensional MHD simulations of accretion disk

and found that accretion disk exhibited strong fluctuations and that intermittently produced strong electromagnetic pulses and matter out of equilibrium was injected toward the rotational axes. The pointing fluxes agreed with those assumed by Ebisuzaki and Tajima [11,12].

On the other hand, the directions of arrival distribution of cosmic rays are accurately isotropic since the discovery cosmic rays. It has been believed that this is because the magnetic field in the galaxy bends the direction of the charged particles tube almost completely isotropic. However, some anisotropy is expected to be observed at energies of $10^{20}$ eV because protons deflect only a few degrees even in the galactic magnetic field [21], and only about 100 Mpc can propagate as a result of collisions with microwave background radiation (GZK-effect [29,30]). In fact, the Telescope Array (TA) team suggested that there is a hotspot in the northern sky where the directions of arrival of the UHECRs [31]. In addition, The Pierre Auger Observatory team also reported that similar hotspots in the southern sky, although it is less statistically significant [32,33]. These are signs of the first anisotropy of charged particles and are key clues to reveal the origin of cosmic rays. He et al. [34] divided the events belonging to the northern hotspot into two by energy, and found that there was a systematic deviation between them. Assuming that this is due to the deflection by the magnetic field, the position of the true source was estimated. The estimated position, though extended to 10 degrees, included several high-energy celestial objects such as M82 and Mrk 180, only M82 was located within the GZK-horizon (~100 Mpc) that the UHECR could reach. M82 is a starburst galaxy and contains many supernova remnants and high-energy objects. It also emits intense gamma rays (1-100 GeV) [35].

In the present paper, we first outline the acceleration theory of wakefield acceleration in the accreting blackhole system (section 2). Next, we apply the theory to the M82 X-1 of ultra-luminous x-ray sources (ULXS) and show that the northern hot spots suggested by the TA experiment can be successfully explained (section 3). In section 4, we discuss the implications from the various astrophysical aspects: those are other nearby active galactic nuclei in the local supercluster, the roles of the intermediate blackholes, such as M82 X-1 in the galaxy formation and evolution, neutron star merging events, and high energy neutrinos. We also discuss the possibility of a similar wakefield acceleration, which accompanies the formation of accretion disk and jets, even in the coalescence process of neutron star detected by gravitational wave detectors. They can clarify the origin of cosmic rays by multi-messenger astronomy as well as neutrino detectors (IceCube and POEMMA), gamma-ray detectors (Fermi gamma-ray astronomy, Cherenkov Telescope Array), visible-infrared ray telescopes, and radio telescopes. In section 5, we conclude the emergence of new mechanism of charged particle acceleration beyond the conventional framework of Fermi acceleration.

2. Bow wakefield acceleration in accreting blackhole systems

Accreting gas forms a disk around a blackhole [36]. In the accretion disk, gas move slowly inward while orbiting in a circular orbit around the blackhole. The orbital velocity and orbital angular velocity are given as follows:

$$v_\varphi = \left(\frac{GM_{BH}}{R}\right)^{\frac{1}{2}} = \frac{c}{\sqrt{6}}\frac{1}{r^{\frac{1}{2}}}, \quad (1)$$

$$\Omega = \left(\frac{GM_{BH}}{R^3}\right)^{1/2} = \frac{c}{\sqrt{6}R_0}\frac{1}{mr^{1/2}}. \quad (2)$$

Here, $m$ is the blackhole mass normalized by solar mass ($M_\odot$) and $r$ is the distance from the center of the blackhole normalized by the radius, $R_{ISCO}$, that of the innermost stable circular orbit (ISCO):

$$R_{ISCO} = 3R_g = \frac{6GmM_\odot}{c^2} = R_0 m, \quad (3)$$

where

$$R_0 = \frac{6GM_\odot}{c^2} \quad (4)$$

is the ISCO radius of a solar mass blackhole. In other words,

$$R = R_0 mr \quad (5)$$

Inside the ISCO, the circular orbits are unstable due to relativistic effects, and the gas falls down at approximately the speed of light and are sucked into the blackholes. In other words, ISCO (r=1) is the innermost radius of the gas disk. In the rest of the section, we will deduce the physical quantities in the disk and jets from the physical constants. We summarize the results and the actual values in Table 1 and 2 for the convenience of the readers.

2.1. Structure of the Steady-state accretion disk

From the mass conservation law, we get:

$$2\pi R\frac{\partial u_D}{\partial t} - \frac{\partial}{\partial R}(2\pi R u_D v_r) = 0. \quad (6)$$

Here,

$$u_D = 2\int_0^{z_D} \rho dz \quad (7)$$

is the surface density of the gas in the disk and $z$ the coordinate in the height direction of the disk. The thickness, $z_D$, of the disk is given by:

$$z_D = (v_s/v_\varphi)R = v_s/\Omega. \quad (8)$$

Next, from the conservation law of angular momentum, we get:

$$R\frac{\partial(2\pi R^2 u_D \Omega)}{\partial t} + \frac{\partial}{\partial R}(2\pi R^3 u_D v_r \Omega) = \frac{\partial}{\partial R}(2\pi R^2 W_{r\varphi}), \quad (9)$$

where $W_{r\varphi}$ is the integral of the frictional stress $w_{r\varphi}$ between adjacent layers in the height direction, in other words:

$$W_{r\varphi} = 2\int_0^{z_D} w_{r\varphi} dz. \quad (10)$$

Since the Keplerian angular velocity is greater at the inside, the frictional forces between adjacent layers cause outward transportation of angular momentum and inward motion of the gas in the disk

In addition, energy conservation law is given by:

$$\frac{\partial \varepsilon}{\partial t} = Q - \frac{4\varepsilon}{3\sigma u_D}, \quad (11)$$

where $\varepsilon$ is the internal energy density and $Q$ the rate of heat generation per unit area in the disk which is given by:

$$Q = \frac{3}{8\pi}\dot{M}\Omega^2. \quad (12)$$

Here, $\sigma$ is the opacity of the gas. According to Shakura and Sunyaev [36], in the innermost area of the accretion disk around the blackhole, the opacity is determined by the electron-scattering process, in other words:

$$\sigma = \sigma_T, \quad (13)$$

where

$$\sigma_T = 0.2(1 + X) \ [cm^2 \ g^{-1}], \quad (14)$$

and $X = 0.7$ is the hydrogen concentration of the gas.

Assuming steady state ($\frac{\partial}{\partial t} = 0$), equations 6, 9, and 11 are reduced to:

$$\frac{d(2\pi R u_D v_r)}{dR} = 0, \quad (15)$$

$$\frac{\partial}{\partial R}(2\pi R^3 u_D v_r \Omega) = \frac{\partial}{\partial R}(2\pi R^2 W_{r\varphi}), \quad (16)$$

$$\varepsilon = \frac{3}{4}\frac{Q}{c}\sigma u_D. \quad (17)$$

By integrating equation 15, we get.

$$\dot{M} = -2\pi R u_D v_r = const., \quad (18)$$

where $\dot{M} = 2\pi R u_D v_r$ is a mass accretion rate, and is constant independent to $R$, the distance to the center of the blackhole. On the other hand, by integrating equation 16, we get:

$$2\pi R^3 u_D v_r \Omega + 2\pi R^2 W_{r\varphi} = const.=0. \quad (19)$$

Here, unlike Shakura and Sunyaev [36], we assume that there were no external torques is imposed at the inner edge ($R = R_0$; ISCO) of the disk. Substituting equation 18 into

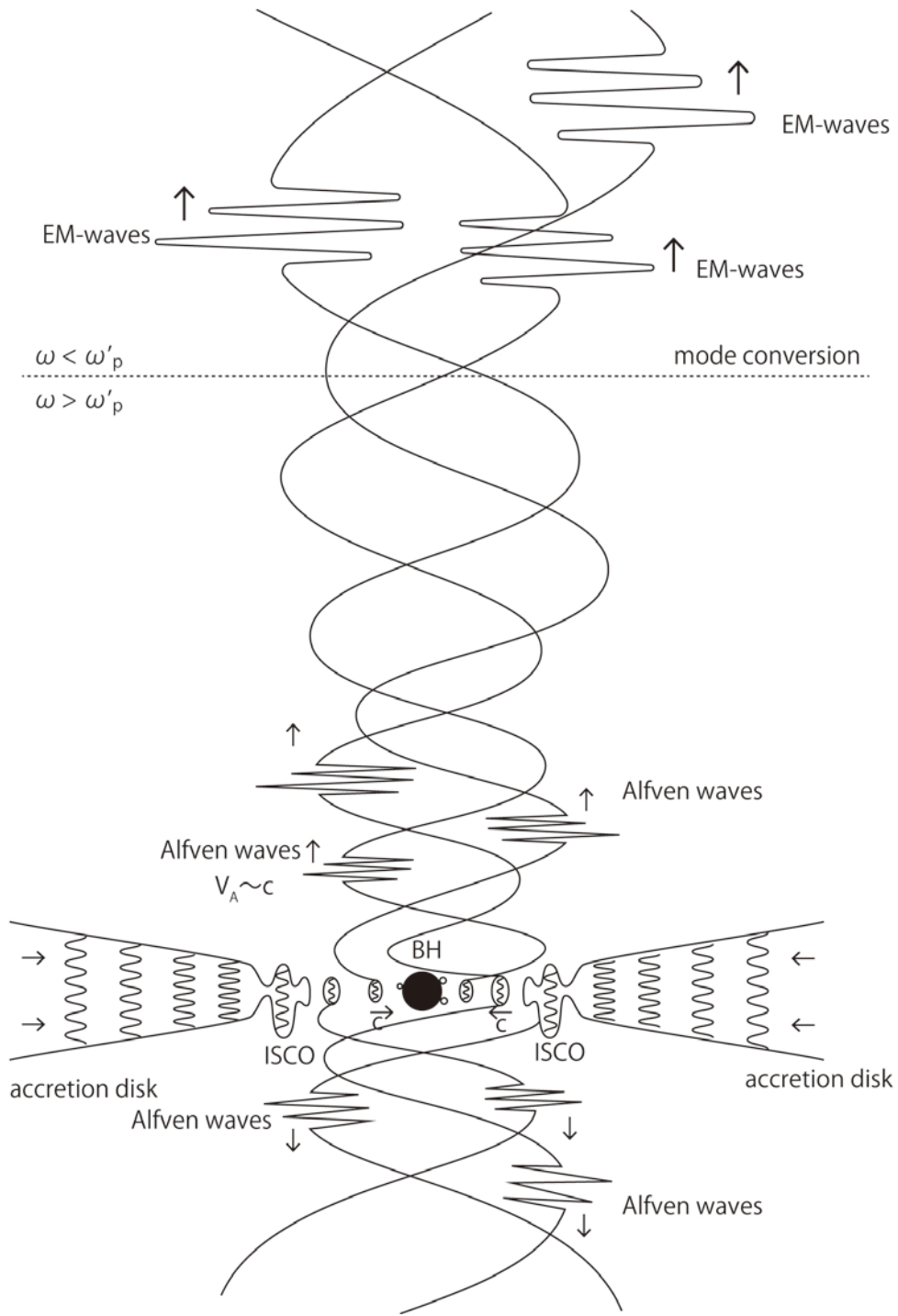

Figure 2  Gas clumps are formed around the inner edge of the accretion disk. When they fall down to the blackhole during its transition, magnetic field penetrating jets are strongly shaken and electro-magnetic disturbances propagates along the jets as bursts of the Alfven/whistler waves.

equation 16, one can get:

$$\dot{M}\Omega = 2\pi W_{r\varphi}. \quad (20)$$

Here, it is assumed that the frictional stress $w_{r\varphi}$ is proportional to the gas pressure $\rho v_s^2$, in other words:

$$w_{r\varphi} = \alpha \rho v_s^2. \quad (21)$$

Here, $\alpha$ is the proportionality coefficient, which is an important parameter defining accretion disk. By substituting equation 21 to equation 10, we get:

$$W_{r\varphi} = \alpha u_D v_s^2. \quad (22)$$

Substituting it into equation 20, we get:

$$\dot{M}\Omega = 2\pi\alpha u_D v_s^2. \quad (23)$$

On the other hand, by substituting equations 12 and 23 into equation 17, we get:

$$\mathcal{E} = \frac{9}{32\pi}\frac{\sigma u_D}{c}\dot{M}\Omega^2 \quad (24)$$

From $v_s^2 = \mathcal{E}/(3\rho)$ and $u_D = 2\rho_D z_D$, we obtain:

$$z_D = \frac{v_s}{v_\varphi} R = \frac{1}{\Omega}\left(\frac{\mathcal{E}}{3\rho}\right)^{1/2} = \frac{1}{\Omega}\left(\frac{2\mathcal{E}}{3 u_D}\right)^{1/2} z_D^{1/2}. \quad (25)$$

Solving this equation for $z_D$ and substituting equations 24 and 14 into 25, we get:

$$z_D = \frac{1}{\Omega^2}\frac{2\mathcal{E}}{3 u_D} = \frac{3\sigma_T}{16\pi c}\dot{M}, \quad (26)$$

Here, $z_D$ is the constant independent to $r$, the distance from the center of blackhole [36].

Radiation luminosity, $L_{\text{rad}}$, is given by:

$$L_{\text{rad}} = \epsilon c^2 \dot{M} = \frac{4\pi c G M_\odot}{\sigma_T}\dot{m}m = \frac{2\pi c^3 R_0}{3\sigma_T}\dot{m}m, \quad (27)$$

where $\epsilon = 0.06$ is the radiation efficiency of the accretion disk [36]. On the other hand, mass accretion rate $\dot{M}$ is given by:

$$\dot{M} = \dot{M}_c \dot{m}, \quad (28)$$

where $\dot{M}_c$ is the critical accretion rate. Here, $\dot{m}$ is the accretion rate normalized by the critical accretion rate $\dot{M}_c$:

$$\dot{M}_c = \frac{L_{\text{Edd}}}{\epsilon c^2} = \frac{4\pi G M_\odot}{c\epsilon\sigma_T}m = \frac{2\pi c R_0}{3\epsilon\sigma_T}m. \quad (29)$$

Substituting equations 8 and 26 into equation 23, we get the surface density $u_D$:

$$u_D = \frac{\dot{M}\Omega}{2\pi\alpha v_s^2} = \frac{\dot{M}\Omega}{2\pi\alpha(z_D\Omega)^2} = \frac{128\pi c^2}{9\alpha\sigma_T^2\Omega\dot{M}} = \frac{64\sqrt{6}\epsilon}{3\sigma_T}\frac{r^{3/2}}{\alpha\dot{m}}. \quad (30)$$

Using this, the internal energy, $\varepsilon_D$, plasma density, $\rho_D$, and magnetic field strength, $B_D$, in the disk can be calculated as follows:

$$\varepsilon_D = \frac{4c\Omega}{\sigma_T\alpha} = \frac{4c^2}{\sqrt{6}\sigma_T R_0}\frac{1}{\alpha m r^{3/2}}, \quad (31)$$

$$\rho_{\rm D} = \frac{u_{\rm D}}{2z_0} = \frac{1024\pi^2 c^3}{27\sigma_{\rm T}^3 \alpha \Omega \dot{M}^2} = \frac{256\sqrt{6}\epsilon^2}{3\sigma_{\rm T} R_0} \frac{r^{3/2}}{\alpha \dot{m}^2\, m}, \quad (32)$$

$$B_{\rm D} = \left(\frac{4\pi}{3}\alpha\varepsilon_{\rm D}\right)^{1/2}. \quad (33)$$

2.2.  Burst emission of the electro-magnetic waves

According to Shibata et al. [26], accretion disk shows repeated transitions between a strongly magnetized state and a weekly magnetized state. When transitioning from a strongly magnetized state to a weakly magnetized state, bursts of electromagnetic waves are emitted [11,12].

The wavelengths of the emitted electro-magnetic disturbances are of the order of the size of the density clamps made in the disk. These are at the wavelength of the most unstable in magneto-rotational instability [37], in other words,

$$\lambda = \left(\frac{V_{\rm A}}{v_{\rm s}}\right)\left(\frac{\Omega}{A}\right) z_{\rm D} = \frac{\sigma_{\rm T} \alpha^{1/2} \dot{M}}{4\pi c} = \frac{R_0}{6\epsilon}\alpha^{1/2}\dot{m}m. \quad (34)$$

Note that, this value is a constant independent to $r$. Here, $V_{\rm AD}$ is the Alfven velocity in the disk:

$$V_{\rm AD} = \frac{B_{\rm D}}{\sqrt{4\pi\rho_{\rm D}}} = \frac{c}{8\sqrt{6}\epsilon}\frac{\alpha^{1/2}\dot{m}}{r^{3/2}}, \quad (35)$$

and $v_{\rm s}$ is the sound velocity:

Table 1 Disk Quantities

| quantities | scaling law | units | equation |
|---|---|---|---|
| $v_\varphi$ | $1.2 \times 10^{10}\ r^{-1/2}$ | cm s$^{-1}$ | 1 |
| $\Omega$ | $1.4 \times 10^4\ m^{-1}r^{-3/2}$ | s$^{-1}$ | 2 |
| $R$ | $8.9 \times 10^5\ mr$ | cm | 5 |
| $z_{\rm D}$ | $1.9 \times 10^6\ \dot{m}m$ | cm | 26 |
| $\dot{M}_c$ | $4.3 \times 10^{-8}\ m$ | M$_\odot$ yr$^{-1}$ | 29 |
| $u_{\rm D}$ | $9.2\ \alpha^{-1}\dot{m}^{-1}r^{3/2}$ | g cm$^{-2}$ | 30 |
| $\varepsilon_{\rm D}$ | $4.9 \times 10^{15}\ \alpha^{-1}m^{-1}r^{-3/2}$ | erg cm$^{-3}$ | 31 |
| $\rho_{\rm D}$ | $2.5 \times 10^{-6}\ \alpha^{-1}\dot{m}^{-2}\,m^{-1}\,r^{3/2}$ | g cm$^{-3}$ | 32 |
| $B_{\rm D}$ | $1.43 \times 10^8\ m^{-1/2}r^{-3/4}$ | G | 33 |
| $\lambda$ | $2.5 \times 10^6\ \alpha^{1/2}\dot{m}m$ | cm | 34 |
| $V_{\rm AD}$ | $2.6 \times 10^{10}\ \alpha^{1/2}\dot{m}r^{-3/2}$ | cm s$^{-1}$ | 35 |
| $v_{\rm s}$ | $2.6 \times 10^{10}\ \dot{m}r^{-3/2}$ | cm s$^{-1}$ | 36 |

$$v_{\rm s} = \sqrt{\frac{\varepsilon_{\rm D}}{3\rho_{\rm D}}} = \frac{c}{8\sqrt{2}\epsilon}\frac{\dot{m}}{r^{3/2}}. \qquad (36)$$

Since the gas in the disk undergoes Keplerian rotation, $\frac{\Omega}{A} = 4/3$. The frequency of electromagnetic wave bursts is given by:

$$\nu = \frac{V_{\rm AD}}{z_{\rm D}} = \frac{c}{\sqrt{6}R_0}\frac{\alpha^{1/2}}{mr^{3/2}}. \qquad (37)$$

On the other hand, the flux of electromagnetic burst, propagating along the jet (Figure 1) is estimated as:

$$\Phi_{\rm w}(r) = \frac{V_{\rm AD}B_{\rm D}{}^2}{4\pi} = \frac{\alpha^{1/2}\omega^2 \dot{M}}{4\pi} = \frac{c^3}{36\epsilon\sigma_{\rm T}R_0}\frac{\alpha^{1/2}\dot{m}}{mr^3}. \qquad (38)$$

In Ebisuzaki and Tajima [11], the wave luminosity, $L_{\rm w}$, of the disturbances electromagnetic disturbances was estimated at $r = 10$. However, Mizuta et al. [28] conducted MHD calculations and found that the wave fluxes are dominated by the innermost regions (ISCO). Therefore, in the present paper, we include the contribution of the wave fluxes from the whole disk in other words,

$$L_{\rm w} = \int_R^\infty 2\Phi_{\rm A}(r)2\pi RdR = 4\pi \int_R^\infty \frac{1}{R^2}dR = \frac{\pi c^3 R_0}{9\epsilon\sigma_{\rm T}}\alpha^{1/2}\dot{m}m. \qquad (39)$$

As a result, the ratio $L_{\rm w}/L_{\rm rad}$ of the wave luminosity to radiation luminosity is as high as unity:

Table 2 Jet Quantities

| quantities | scaling law | units | equation |
|---|---|---|---|
| $n_{\rm J}$ | $1.0 \times 10^9\, \xi\Gamma^{-2}\dot{m}m^{-1}\left(\frac{D}{3R_{\rm g}}\right)^{-1}$ | cm$^{-3}$ | 55 |
| $B_{\rm J}$ | $1.4 \times 10^8\, m^{-1/2}\left(\frac{D}{3R_{\rm g}}\right)^{-1}$ | G | 47 |
| $\omega^*$ | $7.6 \times 10^4\, \alpha^{-1/2}\dot{m}^{-1}m^{-1}$ | Hz | 45 |
| $\omega'_{\rm p}$ | $3.3 \times 10^8 \times \Gamma^{-5/2}\alpha^{-3/8}\xi^{1/2}\dot{m}^{-1/4}m^{-3/4}\left(\frac{D}{3R_{\rm g}}\right)^{-1/4}$ | Hz | 56 |
| $\omega'_{\rm c}$ | $8.3 \times 10^4\, \alpha^{3/4}\dot{m}^{-3/2}m^{-1}\left(\frac{D}{3R_{\rm g}}\right)^{-1/2}$ | Hz | 52 |
| $a_0$ | $3.0 \times 10^{10}\alpha^{3/4}\dot{m}^{3/2}m^{1/2}\left(\frac{D}{3R_{\rm g}}\right)^{-1/2}$ | | 51 |

$$\frac{L_\text{w}}{L_\text{rad}} = \frac{\alpha^{1/2}}{6\epsilon}. \qquad (40)$$

This was around 0.001 in Ebisuzaki and Tajima [11]. At ISCO, wave flux is estimated as:

$$\Phi_\text{w}(r=1) = \frac{\alpha^{1/2}\Omega^2 \dot{M}}{4\pi} = \frac{c^3}{36\epsilon\sigma_\text{T} R_0}\frac{\alpha^{1/2}\dot{m}}{m}. \qquad (41)$$

These waves propagate along the perpendicular (jets) to the accretion disk. Wave flux $\Phi_\text{wJ}(D=3R_\text{g})$ is given by $\Phi_\text{w}(r=1)$. Since the Alfven velocity in the jets is close to the speed of light, electric field $E_\text{w}$ of the wave is calculated as:

$$\Phi_\text{wJ}(D=3R_\text{g}) = \Phi_\text{w}(r=1) = \frac{cE_\text{w}^2}{4\pi}, \qquad (42)$$

where $D$ is the distance from the blackhole along the jet. Therefore, we get:

$$E_\text{w} = \left[\frac{4\pi}{c}\Phi_\text{w}(r=1)\right]^{1/2} = \frac{c}{3}\left(\frac{\pi}{\epsilon\sigma_\text{T} R_0}\right)^{1/2}\frac{\alpha^{1/4}\dot{m}^{1/2}}{m^{1/2}}. \qquad (43)$$

The dimensionless vector potential, $a_0$, at the bottom of the jet is given by:

$$a_0 = \frac{eE_\text{w}}{m_\text{e}\omega_\text{w} c} = \frac{e}{36m_\text{e}c}\left(\frac{R_0}{\pi\epsilon^3\sigma_\text{T}}\right)^{1/2}\alpha^{3/4}\dot{m}^{3/2}m^{1/2}. \qquad (44)$$

Here, the angular frequency $\omega$ of the wave

$$\omega = \frac{2\pi c}{\lambda} = \frac{12\pi\epsilon c}{R_0}\alpha^{-1/2}\dot{m}^{-1}m^{-1}, \qquad (45)$$

where we assume the propagation speed of the wave to be the speed of light, This assumption holds the most of the cases, as can be seen later.

2.3. Wave propagation in the jet

This subsection examines the dependence of physical parameters in the jet on distance from the bottom and discusses how the waves propagate through it. First, the cyclotron frequency $\omega_\text{c}'$ in the jet corrected for relativistic effects is given by:

$$\omega_\text{c}' = \frac{eB_\text{J}}{m_\text{e}c\gamma}. \qquad (46)$$

On the other hand, the magnetic field $B_\text{J}$ in the jet can be calculated assuming that the magnetic field flux is conserved in the jet.

$$B_\text{J} = [B_\text{D}(r=1)](b/3R_\text{g})^{-2} = [B_\text{D}(r=1)]\left(\frac{D}{3R_\text{g}}\right)^{-1} = \left(\frac{16\pi c^2}{3\sqrt{6}\sigma_\text{T} R_0}\right)^{1/2}\frac{1}{m^{1/2}}. \qquad (47)$$

Next, we assume as:

$$\gamma = a_0, \qquad (48)$$

within the jet, $a_0$ can be calculated, assuming that the wave intensity within the jet is conserved, i.e., the flux $\Phi_\text{wJ}$ is inversely proportional to the cross-sectional area $\pi b^2$ of the jet.

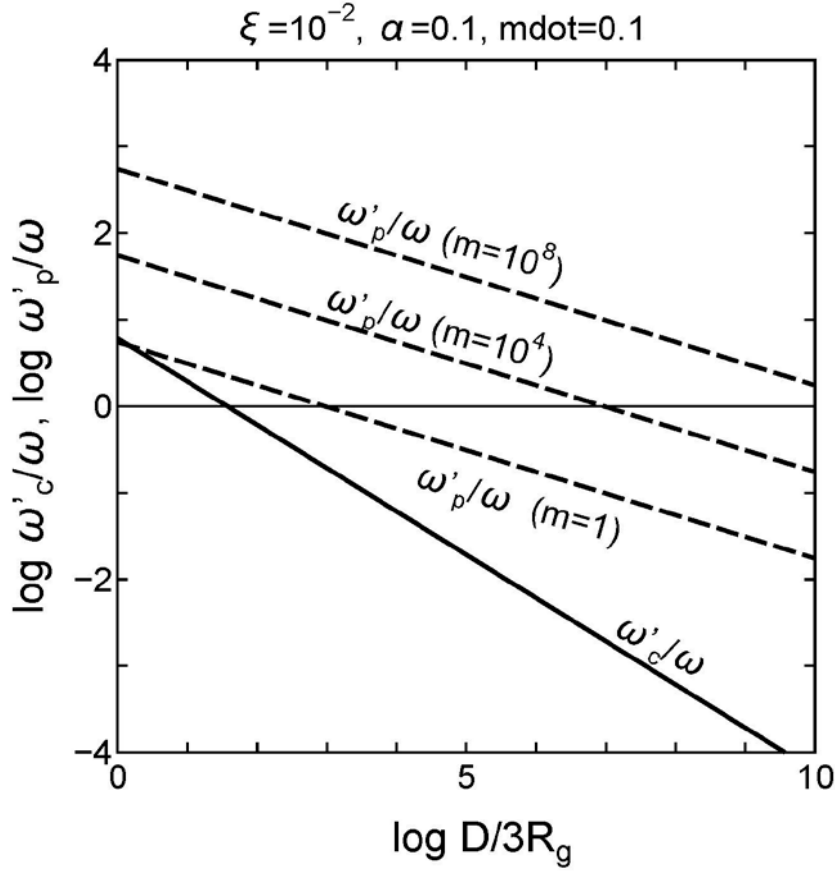

Figure 3 Variation of frequency ratios ($\omega'_c/\omega$: solid line) and ($\omega'_p/\omega$: dashed line) are plotted against the scaled distance $\left(\frac{D}{3R_g}\right)$ for the typical case of $\Gamma = 10$, $\alpha = 0.1$, $\xi = 10^{-2}$, $\dot{m} = 0.1$ for $m = 1, 10^4, 10^8$.

$$a_0(D) = a_0(D = 3R_g) \left(\frac{b(D)}{R_0 m}\right)^{-1}, \quad (49)$$

Where $D$ is the distance from the bottom of the jet, and $b(D)$ is the radius of the jet, which is assumed to $b(0) = 3R_g = R_0 m$. In addition, Figure 2 shows the ratio $\omega'_c/\omega$ of the cyclotron frequency to the wave frequency and that of plasma frequency ($\omega'_p/\omega$) are plotted against the distance $D/3R_g$ from the bottom of the jet for the typical cases ($\Gamma = 10$, $\alpha = 0.1$, $\xi = 10^{-2}$, $\dot{m} = 0.1$ for $m = 1, 10^4, 10^8$). Here we assume that
$$b(D) = R_0 m (D/3R_g)^{1/2} \quad (50)$$
This relation is consistent with the observation of the jet of M87, the closest active galactic nuclei M87 [38]. Therefore, we get:

$$a_0(D) = \frac{e}{36 m_e c} \sqrt{\frac{R_0}{\pi \epsilon^3 \sigma_T}} \alpha^{3/4} \dot{m}^{3/2} m^{1/2} \left(\frac{D}{3R_g}\right)^{-1/2}. \quad (51)$$

Substituting equations 47, 48, and 51 into equation 46, we get:

$$\omega'_c = \frac{144c\pi}{R_0}\left(\frac{\epsilon^3}{3\sqrt{6}}\right)^{1/2}\frac{1}{\alpha^{3/4}\dot{m}^{3/2}m}\left(\frac{D}{3R_g}\right)^{-1/2} = 8.3\times 10^4\ [\text{Hz}]\ \frac{1}{\alpha^{3/4}\dot{m}^{3/2}m}\left(\frac{D}{3R_g}\right)^{-1/2}. \quad (52)$$

On the other hand, the plasma frequency $\omega_p'$ corrected for relativistic effects is given by:

$$\omega'_p = \left(\frac{4\pi n_J e^2}{m_e\gamma\Gamma^3}\right)^{1/2}. \quad (53)$$

The plasma density $n_J$ in the jet can be calculated from as follows, if we assume the kinetic luminosity of the jet:

$$L_J = n_J\mu m_H c^3 \Gamma^2 \pi b^2 = \xi L_{\text{rad}} \quad (54)$$

is conserved through the jet.

$$n_J = \frac{2}{3\mu m_H \sigma_T R_0}\frac{\xi\dot{m}}{\Gamma^2 m}\left(\frac{D}{R_0}\right)^{-1}. \quad (55)$$

Here, $\xi$ is the ratio of the kinetic luminosity of the jet to the radiation luminosity, $\Gamma$ the bulk Lorentz factor, and $\mu$ is the mean molecular weight of the accreting gas. Substituting equations 55, 48, and 58 into equation 53, we get:

$$\omega'_p = \left(\frac{4\pi n_J e^2}{m_e\gamma\Gamma^3}\right)^{1/2} = \left(\frac{96\pi ec}{\mu m_H}\right)^{1/2}\left(\frac{\pi\epsilon^3}{R_0^3 \sigma_T}\right)^{1/4}\frac{\xi^{1/2}}{\Gamma^{5/2}\alpha^{3/8}\dot{m}^{1/4}m^{3/4}}\left(\frac{D}{3R_g}\right)^{-1/4}. \quad (56)$$

Table 3 Time Scales, Maximum Energy, and Luminosities

| quantities | scaling law | units | equation |
|---|---|---|---|
| $2\pi/\omega^*$ | $8.3\times 10^{-5}\ \alpha^{1/2}\dot{m}m$ | s | |
| $1/\nu$ | $7.2\times 10^{-5}\ \alpha^{-1/2}\ m$ | s | |
| $W_{\max}$ | $2.7\times 10^8$ [erg] $z\Gamma\alpha^{2/3}\dot{m}^{4/3}m^{2/3}$ | erg | 65 |
| | $1.6\times 10^{20}$ [eV] $z\Gamma\alpha^{2/3}\dot{m}^{4/3}m^{2/3}$ | eV | |
| $L_{\text{rad}}$ | $1.5\times 10^{38}\ \dot{m}m$ | erg s$^{-1}$ | 27 |
| $L_w$ | $4.1\times 10^{38}\ \alpha^{1/2}\dot{m}m$ | erg s$^{-1}$ | 39 |
| $L_{\text{TCR}}$ | $8.7\times 10^{-2}\left(\frac{\kappa}{0.1}\right)\left(\frac{\alpha}{0.1}\right)^{\frac{1}{2}}L_{\text{rad}}$ | erg s$^{-1}$ | 71 |
| $L_{\text{UHECR}}$ | $8.7\times 10^{-3}\left(\frac{\kappa\zeta}{0.01}\right)\left(\frac{\alpha}{0.1}\right)^{\frac{1}{2}}L_{\text{rad}}$ | erg s$^{-1}$ | 70 |
| $L_{T\gamma}$ | $8.7\times 10^{-2}\left(\frac{\kappa}{0.1}\right)\left(\frac{\alpha}{0.1}\right)^{\frac{1}{2}}L_{\text{rad}}$ | erg s$^{-1}$ | 73 |

In Figure 3, we plot $\omega_c'$, $\omega_p'$ and $\omega^*$ at the bottom of the jet ($D = 3R_g$) against the blackhole mass $m$ for the typical case ($\Gamma = 10$、$\alpha = 0.1$、$\xi = 10^{-2}$、$\dot{m} = 0.1$). For most of the interesting cases, the relationship of $\omega_c'$, $\omega_p' > \omega$ holds; In other words, at the bottom of the jets, the plasma in the overdense state ($\omega_p' > \omega$), where plasma waves and electromagnetic waves cannot propagate. On the other hand, Alfven wave or whistler wave can propagate, since $\omega_c' > \omega$, the Alfven velocity $V_{AJ}$ at the bottom of the jet are given by:

$$V_{AJ} = \frac{B_J}{\sqrt{4\pi m_H n_J}} = \left(\frac{2}{\sqrt{6}}\right)^{\frac{1}{2}} c \frac{\Gamma}{\xi^{\frac{1}{2}} \dot{m}^{\frac{1}{2}}}. \qquad (57)$$

In other words, the nominal values of the Alfven velocity,

$$V_{AJ} \sim 10^{12} \text{ [cm s}^{-1}\text{]} \left(\frac{\Gamma}{10}\right)\left(\frac{\xi}{10^{-2}}\right)^{-\frac{1}{2}}. \qquad (58)$$

This can approach the speed of light, when the approximation breaks down. Then the wave becomes that of EM waves in magnetized plasma.

On the other hand, $\omega_p' = \omega$ at the distance $D_2$ given by:

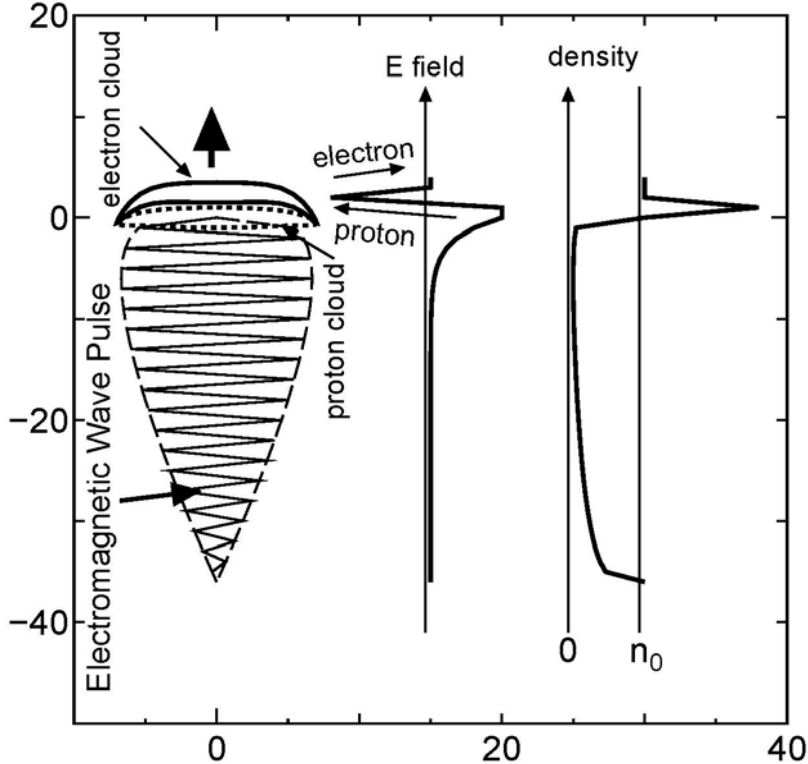

Figure 4. The structure of bow wake. An electron cloud is formed at the front top of the wave pule and a proton cloud follows. The resultant electric field accelerates protons in the back side and electrons in front side of the bow wake (taken from [11]).

$$\left(\frac{D_2}{R_0}\right) = \frac{4R_0 e^2}{9\pi\mu^2 m_H^2 c^2 \epsilon} \frac{\xi^2 \alpha^{1/2} \dot{m}^3 m}{\Gamma^{10}}. \tag{59}$$

On the outside of the point $D_2$ ($D > D_2$), $\omega > \omega'_p$ so that the plasma wave (electromagnetic wave) allow to propagate. The electromagnetic waves propagated as Alfven wave and whistler wave are converted into plasma waves (electromagnetic waves) by nonlinear mode-conversion. This $D > D_2$ leads to the bow wakefield acceleration as described in the next subsection.

2.4. Bow wakefield acceleration

The pondermotive force, $F_{pm}$, for the electrons of the electromagnetic wave is a force generated from the Lorentz force, $\left(\frac{v}{c}\right) \times B$, in the propagation direction of the electromagnetic wave. If the motion of the electrons by the wave is not relativistic ($a < 1$), it can be calculated as the force resulting from the average of the profiles of the electro-magnetic pulses [39-41]. In the relative regime ($a > 1$), this force is more simplified. Since the particle velocity approaches to the light velocity and the plasma satisfies the underdense ($\omega > \omega'_p$) condition as well. Since to the particle velocity asymptotically approaches to the light velocity and that the plasma satisfies the underdense ($\omega > \omega'_p$) condition as well, B is equal to E, in other words, $B = E$. In this case, $F_{pm}$, is given by:

$$F_{pm} = \Gamma m_e eca\omega_w. \tag{60}$$

Charged particles are accelerated by an electric field generated by bow wakefield (longitudinal polarization of electronic distributions). As shown in the Figure 4, protons are accelerated at the back slope of the wakefield, while electrons are accelerated at the front slope. The acceleration force $F_{acc}$ is given by:

$$F_{acc} = z\,F_{pm} = z\Gamma e E_W \left(\frac{D}{R_0}\right)^{-1/2} = \frac{ec}{3}\left(\frac{\pi}{\epsilon\sigma_T R_0}\right)^{1/2} \frac{z\Gamma\alpha^{1/4}\dot{m}^{1/2}}{m^{1/2}} \left(\frac{D}{3R_g}\right)^{-1/2}. \tag{61}$$

Here, $z$ is the number of charges in the particle. The maximum energy, $W_{max}$, obtained by the particle is determined by integrating $F_{acc}$ to the acceleration distance, $D_3$.

$$W_{max} = \int_0^{D_3} F_{acc} dD = \frac{ec}{3}\left(\frac{\pi}{\epsilon\sigma_T R_0}\right)^{1/2} \frac{z\Gamma\alpha^{1/4}\dot{m}^{1/2}}{m^{1/2}} \int_0^{D_3} \left(\frac{D}{3R_g}\right)^{-1/2} dD = \frac{2ec}{3}\left(\frac{\pi R_0}{\epsilon\sigma_T}\right)^{1/2} z\Gamma\alpha^{1/4}\dot{m}^{1/2} m^{1/2}. \tag{62}$$

The acceleration distance is evaluated as the distance $D_3$ to which the acceleration distance equals, in other words, it is given by:

$$D_3 = \frac{e}{432 m_e c}\left(\frac{R_0^3}{\pi^3\epsilon^5\sigma_T}\right)^{1/2} \alpha^{5/4}\dot{m}^{5/2} m^{3/2} \left(\frac{D_3}{3R_g}\right)^{-1/2}. \tag{63}$$

We can solve equation 63 on $\left(\frac{D_3}{3R_g}\right)$:

$$\left(\frac{D_3}{3R_g}\right) = \left(\frac{e}{432 m_e c}\right)^{2/3} \left(\frac{R_0^3}{\pi \epsilon^5 \sigma_T}\right)^{1/3} \alpha^{5/6} \dot{m}^{5/3} m^{1/3}. \quad (64)$$

Substituting equation 64 into equation 62, we get:

$$W_{max} = \frac{1}{9} \left(\frac{e^4 c^2 R_0^2}{2 m_e \epsilon^4 \sigma_T^2}\right)^{1/3} z\Gamma \alpha^{2/3} \dot{m}^{4/3} m^{2/3}. \quad (65)$$

Here, we can eliminate $\dot{m}$ using equation 27 as:

$$W_{max} = \frac{1}{6} \left(\frac{3 e^4 \sigma_T^2 L_{rad}^4}{4 \pi^4 m_e c^{10} R_0^2 \epsilon^4}\right)^{1/3} z\Gamma \alpha^{2/3} m^{-2/3}. \quad (66)$$

When such strong acceleration occurs, the energy spectrum $f(W)$ of the charged particles has a power of the exponent $-2$ [19,20], in other words, $f(W) = A(W/W_{min})^{-2}$. This assumption in electron spectrum is consistent with the typical blazer SED (spectrum energy distribution) and variability [42]. Given the energy efficiency, $\kappa$, of charged-particle acceleration, including the conversion of Alfven wave into electromagnetic waves, the total cosmic ray luminosity, $L_{TCR}$, is given by:

$$L_{TCR} = \kappa L_w = \int_{W_{min}}^{W_{max}} W f(W) dW = \frac{A}{W_{min}^2} \int_{W_{min}}^{W_{max}} W^{-1} dW = \frac{A \ln(W_{max}/W_{min})}{W_{min}^2}. \quad (67)$$

In other words,

$$A = \frac{\kappa W_{min}^2}{\ln(W_{max}/W_{min})} L_w. \quad (68)$$

$$L_{UHECR} = \int_{W_0}^{W_{max}} W f(W) dW = \frac{\kappa}{\ln\left(\frac{W_{max}}{W_{min}}\right)} L_w \int_{W_0}^{W_{max}} W^{-1} dW = \frac{\ln(W_{max}/W_0)}{\ln(W_{max}/W_{min})} \kappa L_w = \kappa \zeta L_w \quad . \quad (69)$$

Here,

$$\zeta = \frac{\ln(W_{max}/W_0)}{\ln(W_{max}/W_{min})}. \quad (70)$$

Since the hotspots in northern sky are seen for the events above $5.7 \times 10^{19}$ eV [31], $W_0 = 0.57 \times 10^{20}$ eV. Therefore,

$$L_{\text{UHECR}} = 8.7 \times 10^{-3} \left(\frac{\kappa \zeta}{0.01}\right) \left(\frac{\alpha}{0.1}\right)^{\frac{1}{2}} L_{\text{rad}}. \quad (70)$$

On the other hand, the total luminosity of cosmic rays is estimated as:

$$L_{\text{TCR}} = \kappa L_A = 8.7 \times 10^{-2} \left(\frac{\kappa}{0.1}\right) \left(\frac{\alpha}{0.1}\right)^{\frac{1}{2}} L_{\text{rad}}. \quad (71)$$

In Table 3, we summarize the scaling low for the times scales, maximum energy and luminosities, which will be used in the next section (section 3).

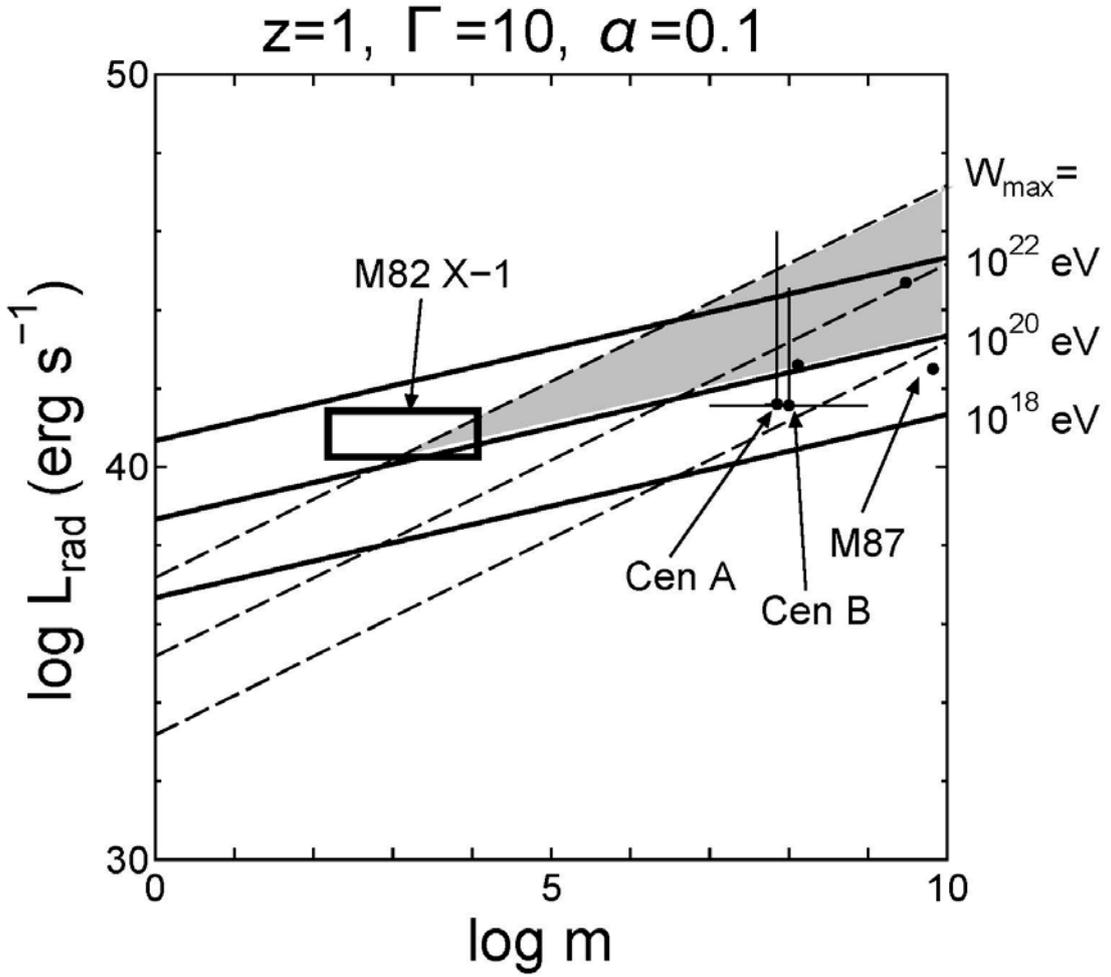

Figure 5. The lines for the maximum energy gain, $W_{\text{max}} = 10^{18}$、$10^{20}$、$10^{22}$ eV, are plotted in $m - L_{\text{rad}}$ diagram. M82 X-1 is located well above the line of $10^{20}$ eV so that be a good candidate for northern hot spot [29]. Other nearby AGNs, such as M87, Cen A, and B are also possible candidates of UHECRs when they are in high luminosity states. Grey area ($W_{\text{max}} > 10^{20}$ eV and $\dot{m} < 0.1$) represents possible acceleration region of UHECR.

3. Starburst galaxy M82

M82 is a nearby (approximately 3.6 Mpc; [43]) edge-on galaxy. M82 is in a starburst state [44] in which a large number of stars are formed at a same time as a result of strong disturbance of the low-temperature gas in the galaxy by the event of the collision with a nearby M81 galaxy. As a result of this starburst, the supernova rate is at least one order of magnitude high compared with a normal galaxy; it is as high as once every 5-10 years [45-47]. In addition, massive and compact young star clusters are seen [48]. Furthermore, many Ultra-Luminous X-ray Sources (ULXS), which have very high luminous intensity ($L_X>10^{39}$ erg/s) been discovered as in other starburst galaxy.

The M82 X-1 is the brightest ULXS in M82, located about 200 pc away from the dynamic center of the galaxy [49-55]. In general, the maximum luminosity of a star bound by gravity is limited by the Eddington luminosity, defined by:

$L_{\text{Edd}} = 1.26 \times 10^{38} \text{ [erg s}^{-1}]m.$ (72)

Thus, it can be concluded that M82 X-1 with an X-ray flux of $10^{40}$ erg/s or more must have a mass of at least 100 $M_\odot$. In addition, the mass of the M82 X-1 estimated from the frequency of the Quasi Periodic Oscillation (QPO) observed in the X-ray luminous intensity as to be 100-1300 $M_\odot$ [50,56,57]. Such a blackhole with mass of $10^2 - 10^5\ M_\odot$ is called an intermediate mass blackhole [58], and is considered to be an important key for solving the mystery of formation of the super massive by connecting a stellar mass blackhole (~10 $M_\odot$) and a supermassive blackhole ($\geq 10^6\ M_\odot$) in a galactic center [59].

3.1. UHECR flux

Applying Ebisuzaki-Tajima theory to the M82 X-1, it has been shown that acceleration to $10^{20}$ eV is well feasible (Figure 5) in the accreting blockhole system. If the distance to the M82 galaxy is 3.6 Mpc = $1.1 \times 10^{25}$ cm, then the UHECR flux, $F_{\text{UHECR}}$, at the Earth is estimated as:

$F_{\text{UHECR}} = \frac{L_{\text{UHECR}}}{4\pi d^2}\left(\frac{4\pi}{2\delta\theta^2}\right) = 1.4\ \left[\frac{\text{UHECRs}}{100\ \text{km}^2\ \text{yr}}\right]\left(\frac{\kappa\zeta}{0.01}\right)\left(\frac{\alpha}{0.1}\right)^{\frac{1}{2}}\left(\frac{\delta\theta}{30°}\right)^{-2}\left(\frac{d}{3.6\ \text{Mpc}}\right)^{-2}\left(\frac{L_{\text{rad}}}{10^{40}\ \text{erg s}^{-1}}\right),$ (73)

where $\delta\theta$ is the UHECR beam divergence angle. On the other hand, the TA team acquired 72 cosmic rays of 57 EeV in 5 years, and observed 19 events (4.5 events of which were expected from uniform arrival) within the hotspot [31]. Since the effective area of TA is 700 km², the observed excess flux in the hot spot direction is about 0.4 UHECRs/100 km²/yr. The observed excess flux in the direction of the hot spot can be explained by the UHECR from the M82 X-1 with the X-ray flux (eq. 65).

3.2. Gamma-ray flux

The Fermi Gamma Ray Observatory detects gamma rays of 0.1-100 GeV from M82. The flux is measured as $(12.2 \pm 1.3) \times 10^{-12}$ erg s$^{-1}$ cm$^{-2}$ [35]. On the other hand, in bow wakefield acceleration, electronics are accelerated as well as protons and atomic nuclei. The energies obtained by electron acceleration is converted into photons (gamma-rays) by some electromagnetic interaction. It's luminosity, $L_\gamma$, is considered to be $L_{\text{TCR}}$ at the highest cases. In other words,

$$L_{T\gamma} \sim L_{\text{TCR}} = 8.7 \times 10^{-2} \left(\frac{\kappa}{0.1}\right) \left(\frac{\alpha}{0.1}\right)^{\frac{1}{2}} L_{\text{rad}}. \qquad (74)$$

For total gamma flux at the Earth can be calculated as:

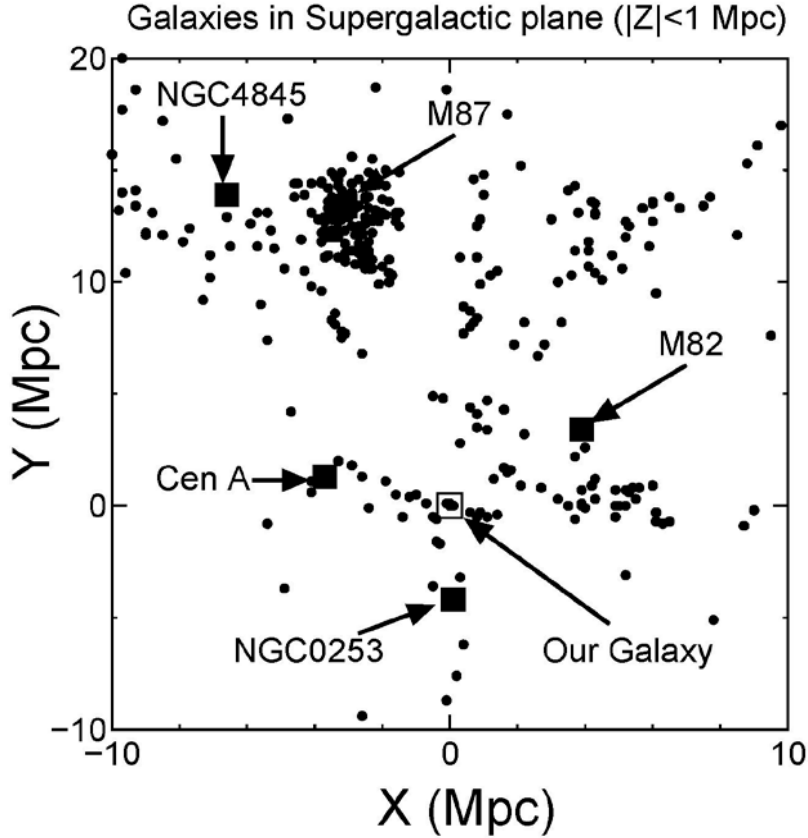

Figure 6. Distributions of galaxies on the long galaxy surface (|Z|<1 Mpc). Our galaxies are represented by open squares. It can be seen that both the heavenly river galaxy and the M82 are in a network of the local large-scale structure. Simulations have shown that a strong magnetic field of the order of 10 nG exists in such mesh networks [57]. Data of the galaxies were taken from https://www.cfa.harvard.edu/~dfabricant/huchra/seminar/lsc/lsc.dat

$$F_{\text{T}\gamma} \cong 4.1 \times 10^{-11} \, [\text{erg s}^{-1} \, \text{cm}^{-2}] \left(\frac{\kappa}{0.1}\right) \left(\frac{\alpha}{0.1}\right)^{\frac{1}{2}} \left(\frac{\delta\theta}{30°}\right)^{-2} \left(\frac{d}{3.6 \, \text{Mpc}}\right)^{-2} \left(\frac{L_{\text{rad}}}{10^{40} \, \text{erg s}^{-1}}\right). \tag{75}$$

The gamma ray flux in the 0.1-100 GeV is consistent with the observed value of the Fermi-Gamma Ray Observatory, assuming that it is several percent or less.

3.3. Deflection due to the magnetic field

If the hot spot in the northern sky comes from M82, the total deflecting angle is 17.4 degrees [34]. If it is due to the deflection by intergalactic magnetic field, we get:

$$\theta = 0.5°z \left(\frac{d}{\text{Mpc}}\right) \left(\frac{B}{\text{nG}}\right), \tag{76}$$

which takes the value of 17.4°. M 82 ($d$=3.6 Mpc) is the source of a proton ($z$=1), each requires a magnetic field of the order of B=9.7 nG. In ordinal intergalactic space, the magnetic field strength is expected as low as 0.1 nG [60]. Therefore, the large deflection angle given by equation 76 cannot be explained.

However, Ryu et al. [61] carried out the simulation of the local large-scale structure of the universe and found that the magnetic field of about 10 nG can be expected in the filament structure. The Figure 6 shows the distribution of galaxies close to the supergalactic plain (within ± 1 Mpc). The distribution of galaxies represents the network (filament structure) of the local supergalaxy to which our galaxies belong. It can be seen that our milky way Galaxy and M82 are in the same filament structure. Therefore, the magnetic field of the UHECR propagating path from M82 to the Milky Way Galaxy can be expected to be about 10 nG, higher than that of the ordinary intergalactic space [61]. It is expected that the UHECR propagating through at distances of 3.6 Mpc is deflected by nearly 20 degrees.

4. Astrophysical Implications

In the present paper, we clarify that an accretion disk around a blackhole emits into electromagnetic bursts propagating along jets, and charged particles are accelerated to energies exceeding ZeV ($10^{21}$ eV) in the electric field in a bow wakefield propagating at a velocity very close to the speed of light. In addition to the electromagnetic disturbances around $r = 10$ [11], those of from the vicinity of the inner edge of the disk was considered. It was shown that the wave luminosity, $L_{\text{w}}$, is comparable to the radiation luminosity, $L_{\text{rad}}$. The results are applied to the intermediate blackhole M82 X-1 in starburst galaxy M82. This object can explain the excessive fluxes of hot spots in the northern sky discovered by the Telescope Array-team[31].

In addition, the structure of the local galaxy was examined, and it was shown that since the space connecting the M82 and the Milky Way Galaxy was almost within a filamentary structure of the local supercluster, the magnetic field could be strong enough to explain the deflection of 20 degrees and even in the case that the main component of the chemical composition of the UHECR was protons.

In the rest part of the section, we discuss astrophysical implications, such as other nearby active galactic nuclei (M87 and Cen A and B), roles of intermediate mass blackholes, such as M87 X-1, gravitational wave burst from the merging neutrons star event, and detection of high energy neutrinos.

4.1. Other nearby AGN in the Local Super-cluster (M87 and Cen A and B)

Other nearby AGN in the local supercluster, such as M87, Cen A and B has a capacity to create UHECRs in their high state (Figure 5). However, a branch of the filaments different from that toward M82 extends towards M87 from the Milky Way Galaxy. Therefore, the path of UHECR from M87 to our galaxy is also considered to be within the filament structure. In fact, galaxies in this direction, and our galaxy, is "falling" towards M87 [62]. The distance to M87 is 16.7 Mpc [63], which is about four times larger than that of M82, the deflection due to the magnetic field would exceed 60 degrees and the spot would spread at least several tens of degrees. Thus, any excess flux from M 87, even if it exists, is difficult to detect with current ground-based detectors.

One more filament extends toward the Cen A. The distances are about 3.4-3.8 Mpc [64-66], which is comparable to M82, so that excess flux to that direction can be expected. In fact, the Pierre Auger Observatory teams reported that there may be excessive fluxes towards the Centaurus region [32,33]. However, the contribution from Cen B, which is much far compared to the Cen A, is difficult to form a confined hot spot. The components from M87 and Cen B would contribute background component of UHECRs. The UEUCR flux over the entire sky was estimated as:

$$\overline{\Phi_{\text{UHECR}}} = 6.7 \text{ [particles/(100 km}^2\text{ yr sr)]} l_{\gamma 37}(\kappa \zeta/0.01)(\tau_8/1.5), \qquad (75)$$

which is consistent to the observational flux of UHECR. Here, $l_{\gamma 37}$ is the local gamma-ray luminosity function of blazers (in the unit of $10^{37} \text{ erg s}^{-1}(\text{Mpc})^{-3}$) and $\tau_8$ is the lifetime of the particles in the unit of $10^8$ yrs. For the case of UHECRs, it is determined by GZK process [29,30].

4.2. Intermediate mass blackholes as the building blocks the supermassive blackholes

The ULXS is considered to be an intermediate mass blackhole of 100-10000 M. Intermediate mass blackholes cannot be made by stellar evolution (e.g.,[65]). However, Portegies-Zwart et al. [67,68] numerically showed that in a dense stars clusters, massive stars of tens of solar mass successively fall to the center of the cluster in the timescale of millions of years due to the of frictional effects acting between stars called dynamical friction, and grow to 100-1000 solar masses by later coalescence with each other. It is considered that the ULXS including the M82 X-1 was made in the high-density star clusters in this way.

This is consistent with the fact that starburst galaxies are rich in both dense star cluster and ULXS [70-72] is consistent with the fact that ULXS and super clusters in starburst galaxies. In fact, many ULXS have been found in super clusters.

Also, such super clusters with intermediate mass blackholes fall towards the center of the galaxy over hundreds of millions of years due to dynamical friction in the galaxy. The blackhole components coalesce with each other and grow, eventually, to a central blackhole of about 1 million solar masses, and evolve into the central core of the Sayfert galaxy. Indeed, many of the Seyfert galaxies are often difficult to distinguish from starburst galaxies (e.g. [73]). Conversely, the star ingredient is discharged to the outside of the star clusters by the reaction and tidal action of the sedimentation of the blackhole component. They are believed to form the so-called galaxy bulge [74]. The idea that a galaxy bulge can be a byproduct of the formation of a central blackhole can explain the correlation between the size of the galaxy central blackhole and the size of the galaxy bulge [75-77] is successfully explained.

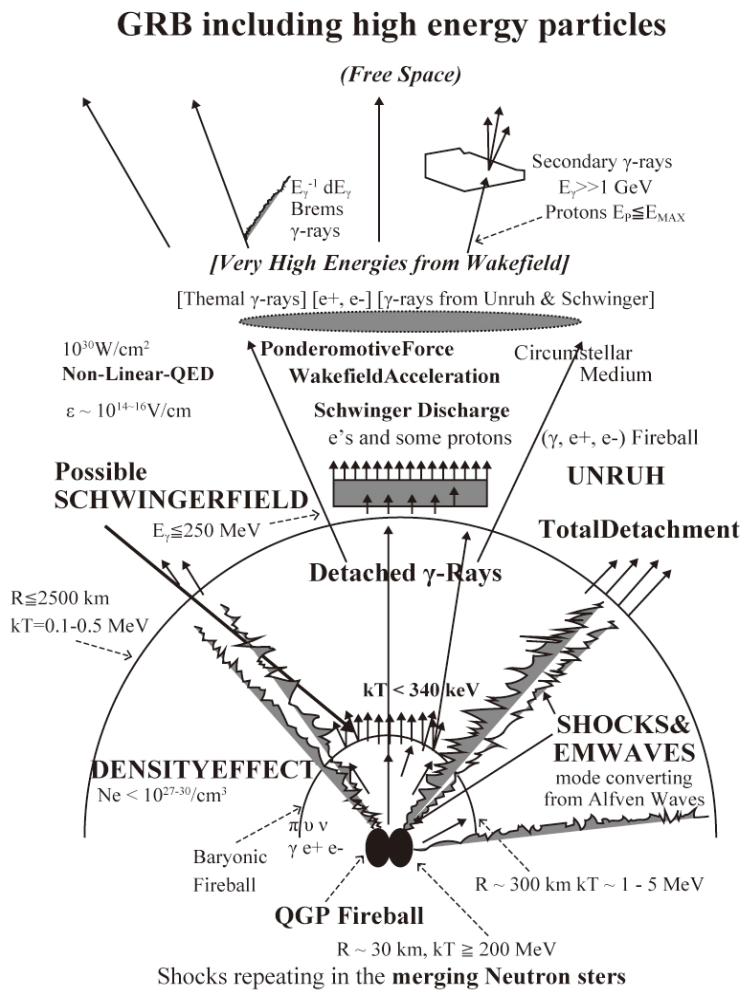

Figure 7. Wakefield acceleration produced by electromagnetic pulses by neutron star merging events are first discussed by Takahashi et al. [23] in terms of quark-gluon plasma (QGP) phase transition (taken from [23] and modified by here).

Because of the presence of a blackhole of $4 \times 10^6\,M_\odot$ in the center of the Milky Way Galaxy, the same process from Starburst to Sayfert Galaxy in the past seems to take place in the Milky Way Galaxy. At present, there is still star clusters in the central region, which are the remnant of the activity, and the presence of the intermediate mass blackhole is suggested in a star cluster [78]. Thus, intermediate mass blackholes such as M82 X-1 and UHECR acceleration are a inevitable process of space evolution of the universe, *i.e.*, the growth of galaxies and the formation of central blackholes.

4.3. Neutron star merging event

Takahashi et al. [23] first discussed neutron star merging event as an origin of UHECRs in terms of quark-gluon plasma phase-transition (figure 7). Takahashi et al. [23] and Chen et al. [20] showed that strong Alfven waves, produced by a neutron star collision were able to accelerate charged particles to energies above $10^{20}$ eV.

On August 17, 2017, for the first time, gravitational waves (GW170817), which seem to be due to neutron-coalescing events, was observed by Advanced LIGO and VIRGO detectors, and at the same time, the Fermi Gamma Ray Observatory detected the GRB170817A of gamma-ray bursts [79]. The latter was about 1.74 seconds behind the former. It demonstrated that the coalescence neutron star was associated with short gamma-ray bursts (e.g., [80]). The neutron stars gradually approach each other with the emission of gravitational waves, and when the lighter stars reach their Roche limits, they are destroyed by the tidal force of the heavier star, to form an accretion disk. This accretion disk becomes a neutrino-cooled accretion disk, NDAF (Neutrino Dominated Accretion Flow [81], rather than electromagnetic wave because of the high temperature (~$10^{9\text{-}10}$ K). Such accretion disk is present for about 1-3 seconds until the heavier neutron star plus the lighter mass is absorbed into the center blackhole of the 2-3 solar mass. If a mechanism similar to that discussed in section 2, $L_w \sim 10^{44-45}$ erg s$^{-1}$ which is 6-7 orders of magnitude higher than that of radiation-cooled accretion disk (section 2), the UHECR can be expected to be formed by acceleration by stronger wakefield. On the other

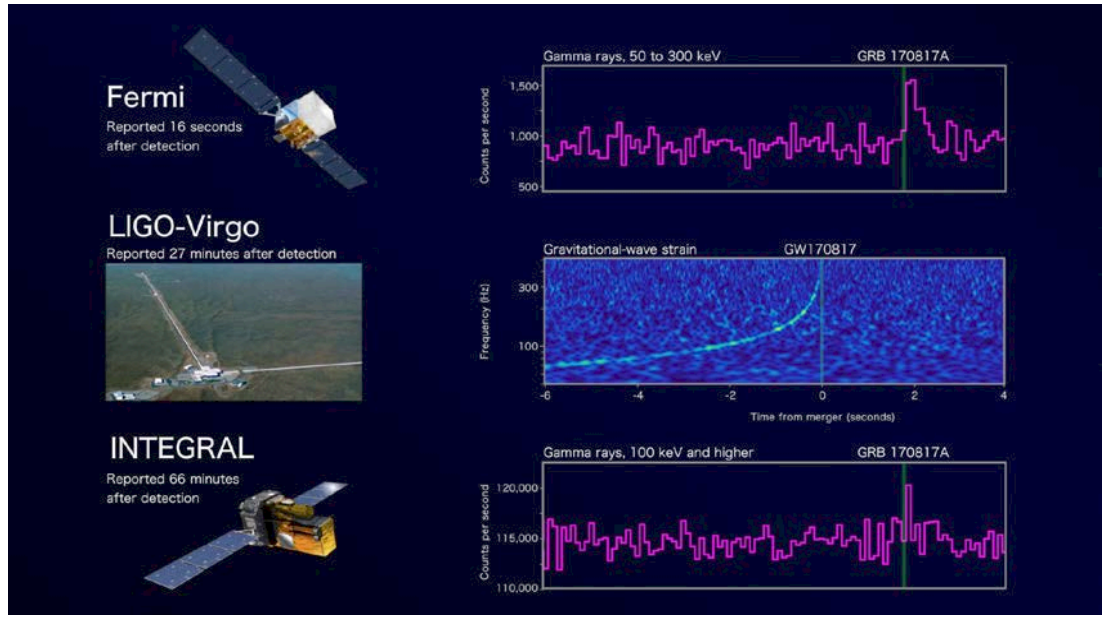

Figure 8. Gamma-ray emission detected by Fermi and Integral satellites from the neutron star merging event (GW178017) delayed by 1.7 seconds compared with gravitational wave burst[79]. This time difference may be explained by the time to build-up the system for the acceleration of charged particles, described in the present paper, in other words, accretion disk and jets. Credit: LIGO; Virgo; Fermi; INTEGRAL; NASA/DOE; NSF; EGO; ESA

(https://heasarc.gsfc.nasa.gov/docs/objects/heapow/archive/transients/gw170817.html)

hand, the coalescence of neutron star should be accompanied by the dispersal of a large amount of materials processed in the star interior. When a beam of accelerated particles enters the materials, ultra-high energy ($10^{20-22}$ eV) neutrinos are emitted by the interaction of the beam and the materials. Since most of the energy used to accelerate electrons is emitted as gamma rays, there is also a possibility that a burst of gamma rays is produced. The delay of the gamma rays to gravitational waves is consistent with the time required for the NDAF disk and jets to form. The formation of the NDAF disk and the formation of the acceleration field due to the coalescence of the neutron stars will be discussed in a separate paper.

4.4. High energy neutrinos

IceCube Collaboration also reported that one 290-TeV neutrino event towards Blazer TXS 0506 + 056 [82]. At that same period, the Blazer was flaring gamma rays (1-400GeV), so it is highly likely that the neutrino came from the object.

At the center of this the object, like the other active galactic nuclei, there is a giant blackhole ($m = 10^{8-9}$). It is thought that the surrounding accretion disk undergoes transitions from a strongly magnetized state to a weak magnetic field state (on a time-scale of 1 month to 1 year), and the associated electromagnetic disturbances propagate through the jets to form a bow wakefield. This bow wakefield accelerates electrons and protons simultaneously. Both have $E^{-2}$ spectra (see [19,20]). The electrons interact with the local magnetic field to emit synchrotron photons, which collide with the electrons to form an inverse Compton peak. That produces, a typical double-peak spectrum of the synchrotron self-Compton emission process [42]. In addition, they have found that the spectral index of 100 GeV gamma rays and the high degree of inverse correlation seen in many Blazars can be explained. At the same time, the accelerated protons become UHECR and collide with the protons in the surrounding material to release neutrinos. Although this object has a red shift of 0.3365±0.0010, and the UHECR does not reach directly, the neutrinos may reach the Earth. The high energy neutrino flux averaged over the entire sky is estimated as:

$$\overline{\Phi_{\text{UHE}\nu}} = 4.7 \times 10^1 \text{ [particles/(100 km}^2 \text{ yr sr)]} l_{\gamma 37} (\kappa \zeta/0.01)(\tau_8/100) \,. \quad (78)$$

For the case of neutrinos, $\tau_8$ is determined by the age of the universe (~15 Gyrs). The bow wakefield acceleration model can explain the IceCube Collaboration observations that flares of gamma rays and neutrino radiations occurred nearly simultaneously, because acceleration of electrons and protons inevitably takes place simultaneously. In the future, the accumulation of similar cases will be awaited.

5. Conclusions

As discussed in the present paper, the bow wake acceleration theory well explains the observational facts related to UHECRs so far. The further verification of the existence of acceleration of ultra-high-energy cosmic rays by bow wake in the universe should be conducted by multi-messenger observations including gravitational wave detectors, such as Advanced LIGO, the VIRGO, and KAGRA [83] and neutrino detectors, such as IceCube [82] and POEMMA [84].

There are two camps of acceleration mechanisms. The first camp is the conventional acceleration of particles by the individual force, including the Fermi stochastic acceleration [15]. The second is the collective force of the plasma through the coherent force ($\propto N^2$, where $N$ is the number of participating particles, as opposed to the individual force [13-15]). Via using the plasma collective acceleration, we find a new path of particle acceleration that may be more intense and more compact that may suit for variety of astrophysical circumstances, some of which have been discussed above.

The origin of ultra-high energy cosmic rays (Ultra-High Energy Cosmic rays: UHECR) with energy of $10^{20}$ eV remain an important puzzle of astronomy, though they are commonly referred to as extragalactic origin, (e.g., [21]). So far, their acceleration mechanisms have been discussed primarily in the framework of the Fermi acceleration mechanism [15]. In the Fermi acceleration mechanisms, it is assumed that charged particles gradually retain energy as they are scattered many times by magnetic clouds. One of the necessary conditions for Fermi acceleration is magnetic confinement [85]. The candidate astronomical objects are neutron stars, active galactic nuclei, gamma-ray bursts, and cosmological accretion shock waves in the intergalactic space. However, even in these candidates, the acceleration to $10^{20}$ eV by the Fermi mechanism has the following problems: (1) very large number of scattering is required to reach high energy; (2) energy loss due to the synchrotron radiation is not negligible at the time of scattering; and (3) adiabatic energy loss takes place when particles escape from the acceleration region. Some of these issues may be addressed by the present mechanism.

Acknowledgements

The work was supported in part by the Norman Rostoker Fund. We would like to thank discussions with Prof. K. Abazajian, Dr. N. Canac, Prof. M. Teshima, Prof. S. Barwick, Dr. A. Mizuta, Prof. B. Barish, Prof. G. Yodh, Prof. H. Sobel, Prof. V. Trimble, Prof. T. Tait, Dr. C.K. Lau, Prof. K. Shibata, Prof. R. Matsumoto, and Prof. H. Sagawa.

81) R. Narayan, T. Piran, P. Kumar, 2001, Accretion Models of Gamma-Ray Bursts, Ap. J., 557, 949-957 (2001).

82) M. Aartsen et al. [The IceCube Collaboration, Fermi-LAT, MAGIC, AGILE, ASAS-SN, HAWC, H.E.S.S., INTEGRAL, Kanata, Kiso, Kapteyn, Liverpool Telescope, Subaru, Swift/NuSTAR, VERITAS, VLA/17B-403 teams], Science. 361, 6398: eaat1378 (2018).

83) T. Akutsu et al., Progress of Theoretical and Experimental Physics, 013F01 (2018).

84) A.V. Olinto et al, ICRC2017, 10-20 July, 2017, Bexco, Busan, Korea (2017).

85) A.M. Hillas, Ann. Rev. Astron. Astrophys. 22, 425, (1984).